# EXPLORING HISTORICAL AND EMERGING PHISHING TECHNIQUES AND MITIGATING THE ASSOCIATED SECURITY RISKS

Marc A. Rader[1] and Syed (Shawon) M. Rahman[2, *]

[1]CapellaUniversity, Minneapolis, MN, USA and Associate Faculty, Cochise CollegeAZ, USA
Mrader3@CapellaUniversity.edu

[2]Associate Professor of Computer Science at the University of Hawaii-Hilo, Hawaii, USA and Part-time Faculty at Capella University, Minneapolis, USA
*SRahman@hawaii.edu

## ABSTRACT

*Organizations invest heavily in technical controls for their Information Assurance (IA) infrastructure. These technical controls mitigate and reduce the risk of damage caused by outsider attacks. Most organizations rely on training to mitigate and reduce risk of non-technical attacks such as social engineering. Organizations lump IA training into small modules that personnel typically rush through because the training programs lack enough depth and creativity to keep a trainee engaged. The key to retaining knowledge is making the information memorable. This paper describes common and emerging attack vectors and how to lower and mitigate the associated risks.*

## 1. INTRODUCTION

Phishing is a social engineering technique that is used to bypass technical controls implemented to mitigate security risks in information systems. People are the weakest link in any security program. Phishing capitalizes on this weakness and exploits human nature in order to gain access to a system or to defraud a person of their assets.

The Anti-phishing Work Group (APWG) is an international group focusing on "eliminating the fraud, crime and identity theft that result from phishing, pharming, malware and email spoofing of all types" [1]. The APWG issues reports semi-annually regarding current trends and emerging attack vectors. The APWG reports that phishing in the second half of 2012 remained at a high level and increased from the first half of 2012. Figure 1 shows the number of phishing sites detected by the APWG for the July through December 2012. This demonstrates a clear threat to organizations and personal information.

Combating phishing requires awareness of phishing attack vectors and methods. This article can be used to improve the content of existing phishing awareness programs that typically target large audiences in a "shot gun" approach to learning where it has a broad spread of information for many targets at once. This approach refines and narrows the subject into a "rifle shot" approach where the audiences contain less people, and the information is more detailed.

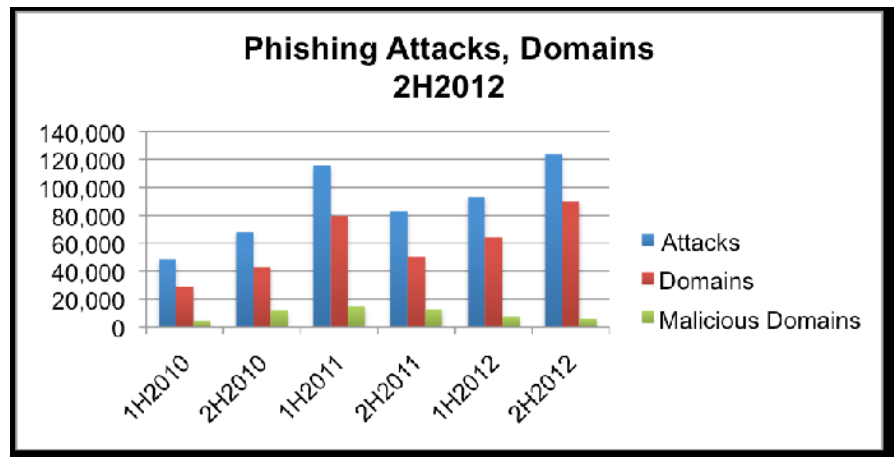

Figure 1 Phishing Sites Detected [1]

The purpose of this article is to provide system administrators and Information Assurance (IA) workforce with a comprehensive awareness program consisting of the history of phishing/social engineering and an exposure to classic and emerging phishing attacks, techniques, and attack vectors. As an organization, much emphasis and funding is spent on implementing and establishing technical controls for Information Security. Information systems are rigorously tested for their vulnerabilities. However, vulnerabilities presented by a Social Engineer deploying a phishing attack cannot be quantified in the same manner since phishing is more of an art than a science and plays on psychology of the target. The investment in technical controls makes it more difficult for an attacker to "hack" into a system. Similar investment must be made in the protections against social engineering or the technical investment is wasted.

> *"Companies spend millions of dollars on firewalls and secure access devices, and its money wasted because none of these measures address the weakest link in the security chain the people who use, administer and operate computer systems"*
>
> --Kevin Mitnick, Ghost in the Wires.

## 2. BACKGROUND STUDY

Dante in his "Devine Comedy: Inferno" devotes the entire $8^{th}$ ring of hell—the Hell of Falsifiers to people who commit fraud. One of the greatest social engineers of all time, Sinon, forever dwells in the "$10^{th}$ bolgia (Italian for Ditch) of the eighth circle of the Inferno" [2]. Although most people do not recognize the name Sinon, people are familiar with his most heinous deed— He was the Greek who convinced the Trojan to bring the Trojan horse inside of the city walls. Sinon is probably the most famous social engineernobody knows. This is the mode of the social engineer; they want to remain anonymous, and preferably undetected. In the case of Troy, the city was fortified and a relentless ten-year Greek siege could not penetrate the Trojan defenses and walls.

In modern times the bricks and rocks of the Trojan fortress are present in our Information Systems. Our critical information systems are guarded by firewalls that are built to keep intruders out. Like Troy, there are ways around firewalls but penetrating a properly configured firewall takes an enormous amount of time, effort, energy and dedication. Like the Greeks, an attacker can test the vulnerabilities of a firewall for months or years and not find any exploitable weaknesses. The easiest way to bypass a network's perimeter security is to convince someone to just let you in. It takes a crafty person like Sinon to convince people to lower their guards even when they are at high alert. This person is known as a social engineer. The best defense against a social engineer is to have a wary and well educated staff [3].

Many organizations have mandatory annual social engineering awareness programs and training. These programs typically cover aspects of social engineering such as: dumpster diving, phishing, masquerading, and others. Typical programs barely touch on the practices of social engineers and do not give enough insight into their motives or methodologies. The result is that there is not a vigilant defense built through training the employees. In order to build a vigilant defense against social engineering the topics need to be broken down and dissected in a way that keeps employees interested so that the training is not just another series of meetings. The most dangerous of the social engineering attack methods is phishing or variants of phishing. The reason that phishing is the most dangerous is the attacker can have complete or near complete anonymity.Phishing is a form of social engineering that is always done remotely, whereas dumpster diving, masquerading, and other social engineering attacks are done face-to-face.

This article focuses on providing information on phishing and its variants in a way that people attending training on anti-phishing techniques will have an interest in the material. This is accomplished by using historical examples, case studies and breaking down the classic and emerging phishing attack vectors. This paper paints a picture of the origins of phishing scams, defines the different types of phishing, how phishing techniques have been used, how they are used currently and prevention.

## 2.1 History of Phishing

The first use of the word Phishing in printed media appeared in an article by Ed Stansel writing for the Florida Times Union and published on March $16^{th}$, 1997. The article stated: "Don't get caught by online 'phishers' angling for account information" [4]. Phishing in the previous sentence is spelled funny because hackers have their own language called Haxor (sometimes synonymous with Leetspeak depending on the source). Typically, Haxor replaces Standard English characters with other ASCII characters. Ypass.net has an English to Haxor translator. An example of a sentence in Haxor is the following:

| English | Fishing can be fun if you have the right bait |
|---|---|
| Haxor: | Phi5HiNg c@n 83 fun iF y0U H@V3 tHe RIgH7 8aI7 |

A typical rule in Haxor is that the letter "f" is converted to "ph". The origin of the word phishing is considered to be an extension to the word "phreaking." "Phreaking was coined by John Draper, aka Captain Crunch, who created the infamous Blue Box that emitted audible tones for hacking telephone systems in the early 1970s" [5]. Phreaking is the hacking of phone networks. Draper was called Captain Crunch because he is credited with discovering a toy whistle included in Captain Crunch cereal that emitted a tone at 2600 Hz that was the exact tone that AT&T services technicians used to make test calls. Blowing the whistle into the phone allowed a person to make free calls. This theft of service transferred into phishing scams.

The earliest phishing scams dealt with stealing a person's credentials to login to an Internet Service Provider (ISP). In the days before broadband internet and free unlimited internet, people accessed the internet via dial-up modem. Once connected to the ISP, a user entered a username and password. ISPs at the time charged users by the minute for their internet usage.

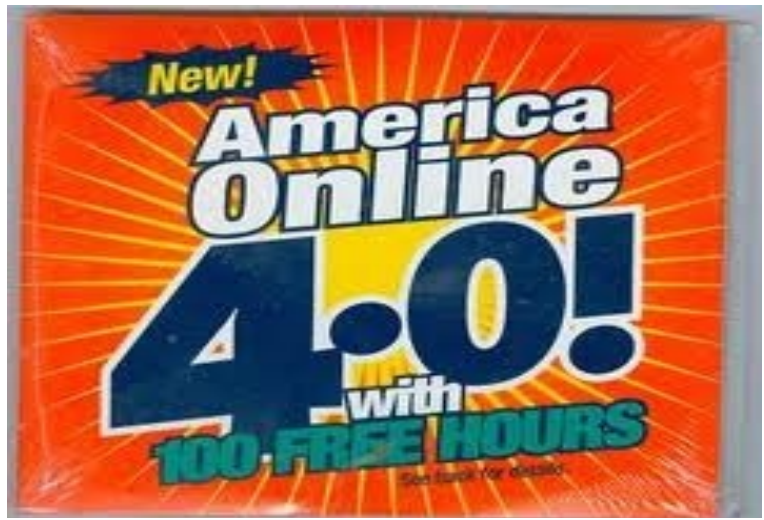

Figure 2. Vintage AOL CD [6]

Above is a picture (Figure 2) of an AOL CD that could be picked up for free at any computer store and most "big box "retailers--note the "100 free hours" that came with the CD. Phishing's origins come from the desire for unlimited internet usage. By compromising an AOL account, a person could have unlimited internet at the expense of the person whose account was compromised. The victim would receive a bill from AOL for the account overage caused by the thief and would have to call AOL to dispute the charges. AOL did directly billed a customer's credit card, so the overage may not have been noticed by the victim for a couple of months.

The most common attack vector for stealing AOL credentials was phishing in AOL chat rooms. A hacker simply known as "Da Chronic" may have earned himself a spot in Dante's eighth ring next to Sinon. Da Chronic developed a program called AOHell. Da Chronic describes the functions of AOHell as:

"Basically, it is used to annoy others, get free service, and other things. You can knock people offline with it, you can Email bomb someone with it, and many other things. You can use it to automatically reply to IMs that you receive, ignore IMs from certain, but not all people, and much more. You can use the Artificial Intelligence bot to greet people who enter the room, and send other messages when certain things happen in the chat room. You can also use the AOHell Fake Account Creator to make fake AOL accounts quickly. You can also use it to get other people's passwords and credit card information" [7]

AOHell was a free tool available to everyone. One of the tools in the toolkit was a program called "Fisher". Fisher was a tool allowing an attacker to masquerade as an AOL administrator and automated the process of creating a chat room that appealed to new users. With the "chat logging" feature turned on, the automated program asked the users for either their user name and/or account information. All the attacker had to do was to search the logs from the chat conversations for the information and compromise either the account or the credit card.

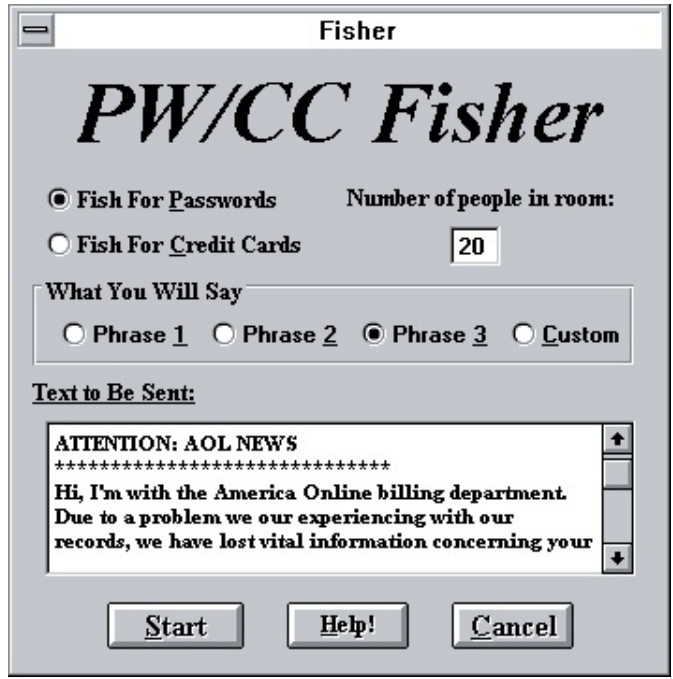

Figure*3*. Fisher Example [8]

AOHell had the ability to exploit AOL's free service offers by generating fictitious credit card numbers for account authentication. "According to users, the program can make free accounts that last up to 10 hours of online time or one week, whichever comes first" [9]. Using AOHell to create a temporary account was a small accomplishment used as a stepping stone to capture a larger prize. Since a new account could only be used for 10 hours, the attacker set his sights on compromising an existing account.

A more sophisticated attack existed in the AOL chat rooms, but it did not involve AOHell. This attack involved malicious websites. The way the attack worked was the attacker created a basic, official looking website. The attacker would initiate conversations in a chat room with a victim and drive the conversation in such a way that the victim was persuaded in leaving AOL chat and going to another website. The attacker directed the user to his own website where the victim was asked to enter his AOL login information into the website. The site recorded the victim's credentials for the attacker. Phishing for AOL accounts was so prevalent in the 1990's that "phish were actually being traded between hackers as a form of currency. Hackers would routinely trade 10 working AOL phish for a piece of hacking software that they needed" [10]

## 3. CLASSIC ATTACK VECTORS

### 3.1 Social Engineering

Social Engineers are experts at appealing to the human psyche. Their most common methods of manipulation rely on Curiosity, Fear, and Empathy. A Social engineer uses these tactics in phishing scams. These tactics are described below:

- **Curiosity.** Exploiting a person's curiosity might involve sending an e-mail that purportedly contains a link to watch a video about the latest sensational news story. The

27

- link, however, will lead to a malicious site aimed at installing malware or stealing private information.
- **Fear.** One tactic cyber thieves use to instill fear and persuade a person to act in a certain way is by sending phishing e-mails, supposedly from a victim's bank. Using the claim that his or her account has been breached, the message will push the user to click a certain link to validate the account. Again, the link will lead to a malicious site aimed at compromising the person's computer, or stealing sensitive information.
- **Empathy.** To take advantage of a person's empathetic feelings towards others, hackers have been known to impersonate victims' friends on networking sites, claiming to urgently need money. In another prime example, recent social engineering scams have also been seen in the wake of the earthquake and tsunami in Japan, with scammers attempting to profit from the tragedy [11].

Curiosity, fear and empathy have been used by the social engineer since the tactics have been recorded. Mentioned earlier, Sinon the Greek was one of the most skilled. Virgil's *The Aeneid* tells the story. After 10 years of fierce battle between the Greeks and Trojans, the Greeks relied on trickery. The story of the Trojan Horse is well known. What is not well known is that the Trojans were convinced by Sinon to bring the horse into the city. Sinon used curiosity, fear, and empathy. When the Greeks left, Sinon and the horse were left behind. Trojans left their walled city and found both Sinon and the horse. Sinon told the Trojans that he was left behind by the Greeks to die at the hands of the Trojans. Sinon condemned the Greeks by emphatically becoming friends with the Trojans. Sinon told the Trojans that the Greeks believed that their failure to destroy Troy was a result of their desecration of a temple devoted to Athena. The Greeks built the horse to honor Athena so that Athena would not disrupt their voyage home in retreat. Sinon convinced the Trojans that if they moved the horse from the Athenian temple and into the walls of Troy that the honor to Athena would be broken and the Greeks would have a cursed voyage home. The rest is commonly known history and the basis for the saying "I fear Greeks even bearing gifts" [12]

### 3.2 Trojaned Hosts

Gifts are a common tactic that attackers use as bait to catch their phish. The ultimate prize of a phisher is a trojaned host. An attacker will send a phishing email that offers the victim some type of gift if the victim visits a website. The gift that the victim receives is a Trojan horse program on their computer. Typically, the Trojan program allows the attacker to have remote control of the computer and access to all information on the computer and information that the computer processes. One of the greatest uses of the compromised computer is propagating the phisher's spam messages.

Unfortunately there is a Trojan creation toolkit available called Zeus (keeping with the mythological theme). Zeus, which is sold on the black market, allowing non-programmers to purchase the technology they need to carry out cybercrimes. According to a 2010 report from SecureWorks, the basic Zeus package starts at about $3,000 [13]. The scary thing about the Zeus package is that a study found "most of the infections occurred on machines where an antivirus product was installed and kept up-to-date: 31% of the Zeus-infected PCs had no antivirus while 55% had updated antivirus software" [14]. Zeus Trojans are typically not detected by anti-virus and is considered a zero-day exploit due to it being a custom tool for creating unique Trojans that are not "in the wild". When the Trojan is created, if the creator limits the use of it, the anti-virus vendors will never add the signature to their databases due to its low distribution.

### 3.3 Man-in-the-middleAttacks

A Man-in-the-Middle (MitM) Attack is the result of an attacker inserting themselves logically into the network between a victim and a legitimate website. A MitM attack employs a number of attack methods such as transparent proxies, DNS Cache Poisoning, URL Obfuscation, and browser proxy configuration attacks.

### 3.3.1 Transparent Proxies

This type of attack uses a traditional phishing methodology of sending a deceptive email to an intended victim. Contained in this email is a spoofed URL for a targeted website. The attacker's intention is to steal personal information. In the figure below the attacker, noted as attacker.org, acts as proxy for the entire transaction. The victim is completely unaware of the proxy in this attack.

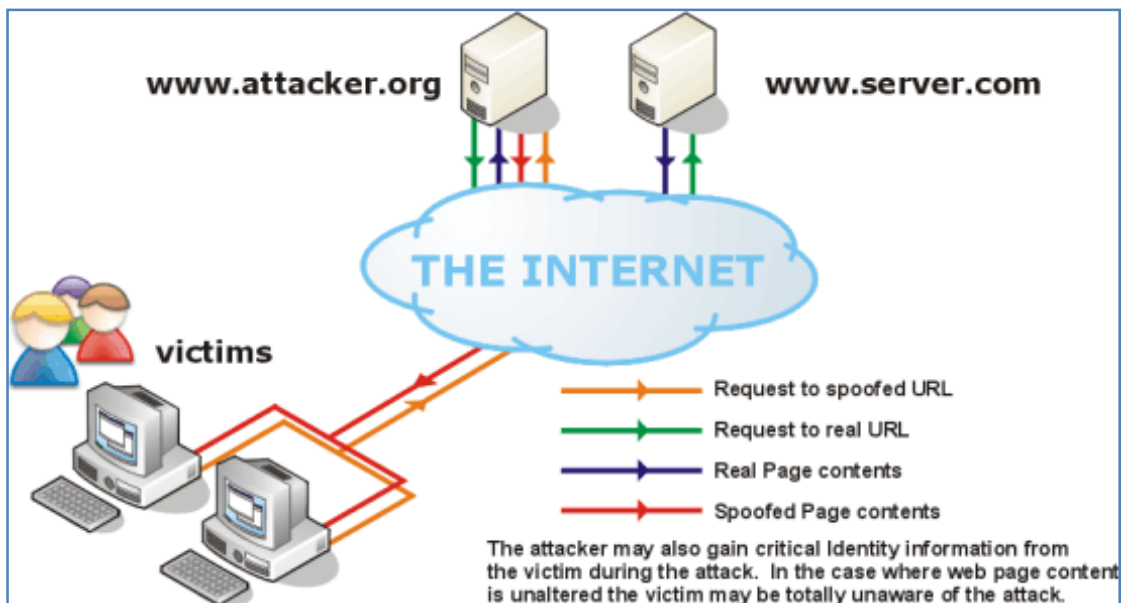

Figure4. Man in the Middle [15]

In this situation, the attacker copies the entire source code from the original website. The attacker modifies the original HTML code and uploads it to the spoofed website. The spoofed website interacts with the source website on behalf of the victim. Following the network lines in the figure, the victim never interacts with the source website directly and the attacker can gain all of the victim's information.

### 3.3.2 DNS Cache Poisoning.

Domain Naming System (DNS) Cache Poisoning is an almost undetectable phishing method. "DNS cache poisoning, is the corruption of an Internet server's domain name system table by replacing an Internet address with that of another, rogue address" [16]. The victim in this case does not respond to an unsolicited email. The victim enters a legitimate web address into a browser. The attacker altered the DNS lookup tables on the DNS server. Web addressing is Internet Protocol (IP) number based. IP addresses in version 4 are based on 4 octets. IP addresses follow this format xxx.xxx.xxx.xxx. Remembering an IP address is difficult for users so DNS was invented so that users can use common names. The computer resolves the IP from a

International Journal of Network Security & Its Applications (IJNSA), Vol.5, No.4, July 2013

DNS server. Resolving means that common names are automatically converted to the IP format. The DNS server has look up tables to make name resolution faster. In DNS poisoning the attacker changes the IP address that is associated with the common name to match a bogus website that the attacker has set up. Figure 5 below outlines the process.

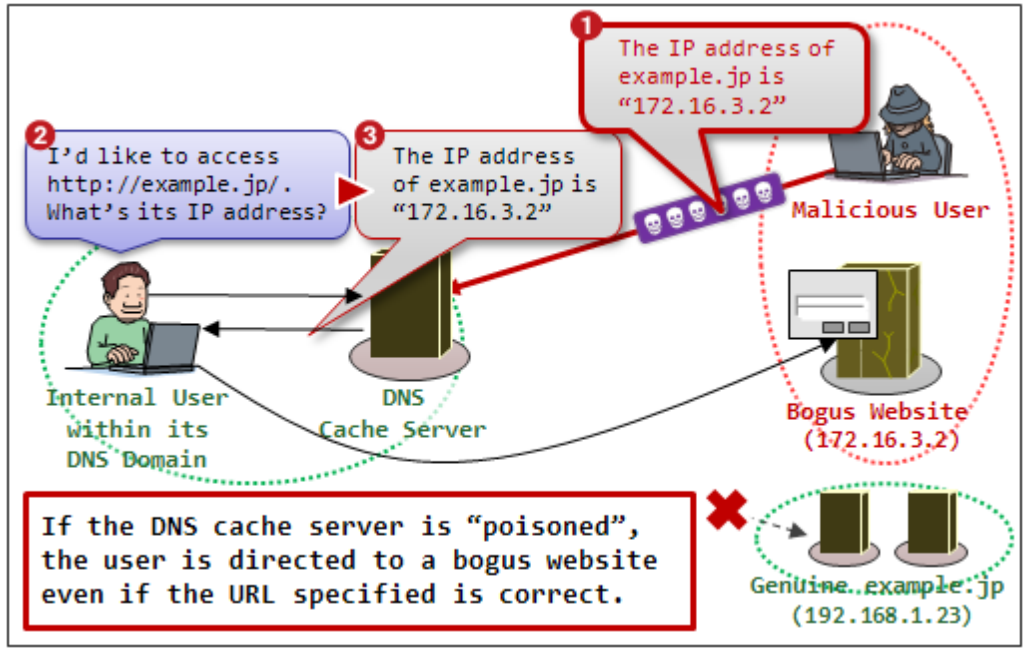

Figure5. DNS Poisoning [17]

### 3.3.3 Browser Proxy Configuration

A Browser Proxy attack is similar to the transparent browser attack since they both use a proxy, however, the browser proxy attack is much easier to detect. In the Browser Proxy attack, configuration changes are made on the victim's proxy setting of their web browser. The attacker stages a server logically between the victim and the internet. The proxy server captures all data that is transferred between the victim and the internet. According to Gunter Ollmann, the phishing part of this attack is typically the second part of the attack. Initially an attacker uses malware to change the proxy information and then a spear phishing email is sent to get the user to enter desired information into a website [18]. Spear phishing is a targeted attack where the attacker spoofs the role of a person that the victim trusts.
One thing that makes MitM attacks so prevalent is that hackers create and release automated tool kits making it easier for hackers to commit crimes, but it also makes it easy for non-technical thieves called "script kiddies" to create and deploy exploits. In 2007 RSA (RSA.org) discovered that hackers were selling a Universal Man-in-the-Middle Phishing kit where an attacker can purchase a kit that automates the creation of the spoofing website. The cost of the kit at the time of publication was approximately one-thousand dollars [19].

### 3.4 URL Obfuscation Attacks

The "bread and butter" of phishing attacks is convincing a victim to click on a link. In the early days of AOL it was easier to convince a user to click on a link or trick them into downloading a payload. Nowadays, computer users are typically more aware of scams and the innocence of the internet has expired. Most users are suspicious of emails. A user is not going to be tricked into clicking on a link for Bank of America with an address of www.I-am-a-badguy.com. Users can read the URL and see that the link is not for Bank of America. Attackers now need to find a way to make their web address look like the real address. Attackers employ many different ways to fake an address; some of these are listed below.

### 3.4.1 Bad Domain Names

Registering a domain normally takes only a few minutes and only costs a couple of dollars. An attacker can obfuscate a URL by registering a domain name that looks like a real domain. One example is registering a domain to look like Microsoft.com. An attacker registers Micros0ft.com where the second "o" is a zero. When spelled out it looks silly and nobody would go to the site, but if the bad domain were in an email listed in all caps it looks like this: MICROS0FT.com. If the attacker manipulated the fonts or the text sizes this domain can look real. In the standard ASCII characters there are characters that look similar such as upper case "i" and a lower case "L" or even a "1". Example: "Il1" is and "i" a "L" and a one. Ollman also mentions that there are characters in other languages that are similar to the English or Standard ASCII characters. He cites the use of the ASCII "o" and the Cyrillic "o" [20].

Another way to obfuscate a domain name is to use a sub-domain. A sub-domain is added to the front of a web address. If Microsoft had a sub domain called "customer service" the domain name would like: www.customerservice.microsoft.com. The employment of sub-domains is easy for an attacker to do since sub-domains are not registered—a domain administrator creates them. If an attacker owns the domain bankfraudalert.com and sets up that domain for phishing, the attacker can make the URL look legitimate such as www.bankofamerica.bankfraudalert.com. Even greater sub-domain obfuscation can make the URL look real by adding a .com to the URL. In this the URL looks like: "www.bankofamerical.com.bankfraudalert.com." The first instance of the .com is actually a sub-domain.

### 3.4.2 Third-party shortened URLs

URLs for legitimate purposes have grown to some unmanageable lengths due to URLs containing multiple sub domains, deep folder structures, and search strings included in the URLs. This creates a need for third party vendors to assist with shortening URLs. The most common are smallurl.com and tinyurl.com. These providers do the opposite of what DNS does, they take a long common language URL and shorten it using characters that cannot be read by a person. Databases on their servers correlate the shortened URL to its common long URL. These services did not intend to create a product to be used by phishers, but the side effects of their product are that they have a great tool for scammers. A scammer does not need to come up with clever ways to spoof a real domain name. The scammer can use the third-party tool for site obfuscation. A good scammer will still use a spoof name, but the third-party obfuscation adds a level of complexity and believability to the scam. In the original use of the spoofed Microsoft domain (Micros0ft.com) the scammer registers the domain with the third party and the domain look similar to this: http://tinyurl.com/7h6kndc. Tiny URL even has an aliasing service that allows a person to change the suffix to whatever they want. The same URL listed above (which was www.microsoft.com/customerservice) was changed to: "http://tinyurl.com/BankofAmericafraudcomplaints." This technique hides the original domain name and appears to be something that it isn't.

### 3.4.3 Host name obfuscation

Computers can be very accommodating with regard to formatting host names. Accommodating the user under the pretext of computer friendliness also makes an attacker's job easier through host name obfuscation. Host names can be entered using DNS, in Decimal, Dword, Hexadecimal and sometimes a mixture of all of them adding a level of complexity and layer of confusion. Technicalinfo.net gives a great example for Host name obfuscation [21]:

These alternative formats are best explained using an example. Consider the URL: http://www.evilsite.com/, resolving to 210.134.161.35. This can be interpreted as:

- Decimal – http://210.134.161.35/
- Dword – http:// 3532038435/
- Octal – http://0322.0206.0241.0043/
- Hexadecimal – http://0xD2.0x86.0xA1.0x23/ or even http://0xD286A123/
- In some cases, it may be possible to mix formats (e.g. http://0322.0x86.161.0043/).

DNS was invented so that users would not have to translate obfuscated host names. Unfortunately it is necessary to allow network to be flexible in its addressing. An attacker can take the mixing of the formats further and make a site look more legitimate such as http://210.134.161.35/BankofAmerica/fraudepartment.

*Note: Even Microsoft Word acknowledged most of the above web addresses as valid while writing this paper before hyperlinks were removed.*

## 3.5 Cross-site Scripting Attacks

Cross-site Scripting (known as CSS and XSS) is a method where an attacker manipulates and exploits insecure coding or poor coding of a website allowing the attacker to inject malicious code, such as a Trojan or key logger, for collection of data. The pie-chart below, created by the Web Hacking Incident Database for 2011 (WHID) shows SQL injection and XSS are the most popular attack vectors on web sites. Adding to this, some of the other attack methods on the chart can be as a result of XSS attacks

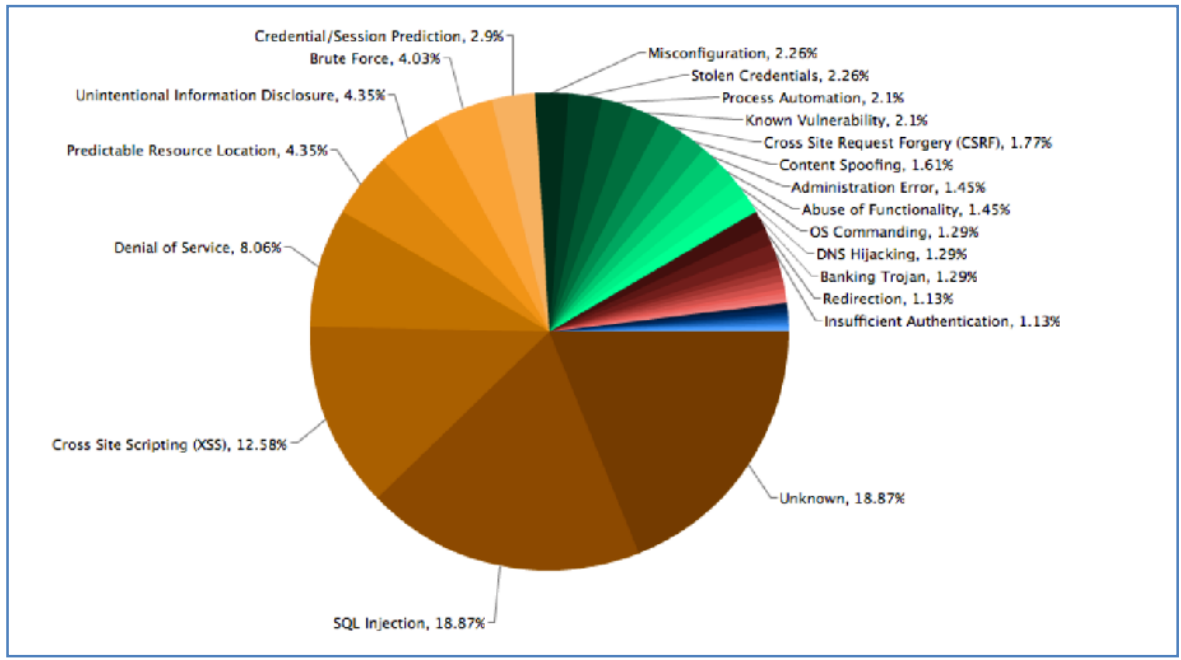

Figure 6. Cross Site Scripting [22]

The most common method used by phishers using XSS isto exploit a website that allows comments on its content. The XSS code is placed into the comments field and auto executes every time the page is loaded by a victim.

## 4. ATTACK BREAKDOWN

### 4.1 Dissecting the Phish

Figure 7 below shows a random representation from an email spam folder. The email shows some tell-tale signs of a Phishing scam.

The original email is sent from CreditScoresNow@Hallcow.com but is showing its name as Equifax. This is not even a good attempt at URL obfuscation by bad domain name. An email from Equifax should come from Equifax.com. The entire body of the text is actually a picture with a hyperlink. Attackers do this to circumvent text scannerslooking for typical phishing attacks. At the bottom of the email there are two links for unsubscribe. The links go to the same server, but different sub-domains. Also, the host names are very long and contain numbers. This is an example of URL Hostname Obfuscation.

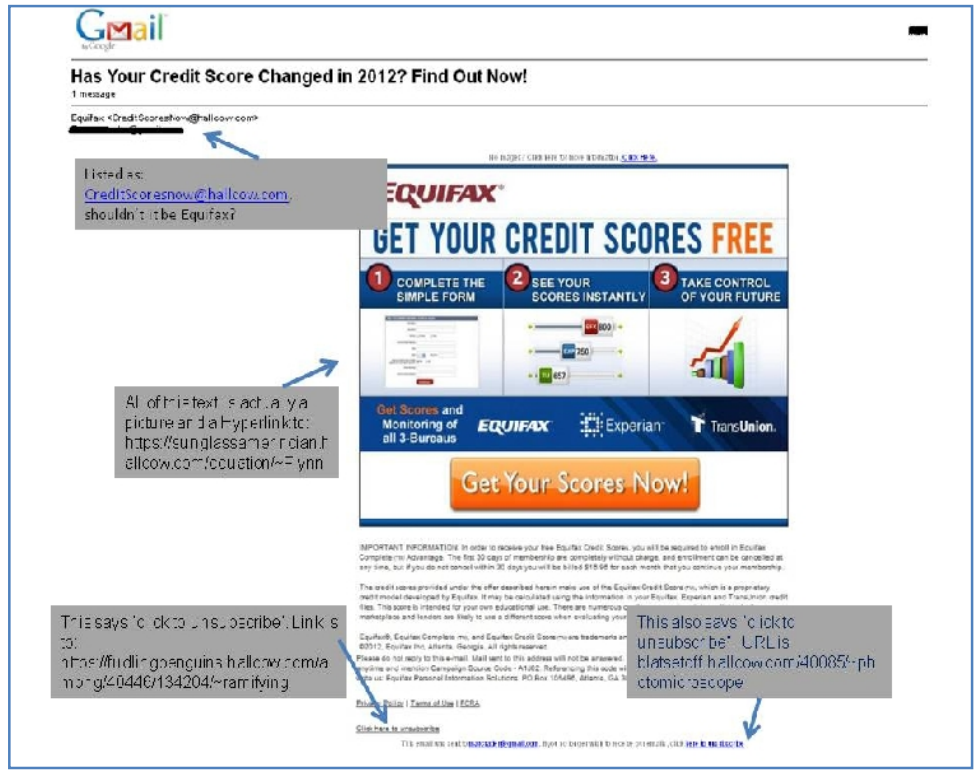

Figure 7 Phishing Example

## 5. EMERGING ATTACK VECTORS

Since the days of AOL's dominance and the immaturity of both the internet and its users, attacks have changed to match internet trends. Users are more aware of classic scams and detection and prevention technologies are developed to counter the methods of the attacker. Attackers deviously modify their techniques to continue to fool both people and technology. Listed below are some emerging attack vectors and new targets of opportunity for phishers.

### 5.1 Social Media

Social media is an excellent source of information for a spear phisher. Facebook and Twitter have become online diaries of people's lives. Diaries used to be secured by lock and key so that little brothers would not read little sister's thoughts. Now these thoughts are broadcast for the entire world to see in real time. Attackers can use this information to construct target phishing scams using information from a person's profile on Facebook. Some people even have their email and phone number posted as part of their profiles giving a scammerinformation needed for email phishing or smishing.

Facebook is also vulnerable to XSS. According to Acunetix, "Something as simple as a Facebook post on your wall can contain a malicious script, which if not filtered by the Facebook servers will be injected into your Wall and execute on the browser of every person who visits your Facebook profile" [23].

### 5.2 SMS

SMS stands for Short Message Service. SMS is basis for cellular phone texting. Cellular phones are now web enabled, and have become targets for an attack called Smishing. Attack vectors present in email messages are now being migrated to text messages. Attackers are using an

34

update of an old technique called war dialing. In the classic sense an attacker used a computer modem to automatically call a list (or all of them in an area code) of phone numbers and record which numbers responded as computers or fax machines. The current scam is to use war dialing to look for cellular numbers and then send out a mass message that directs the recipient to a malicious website.

> **Note:** *Wardialing is also being used for Vishing scams as well.* Below is an example of a Smishing attempt.

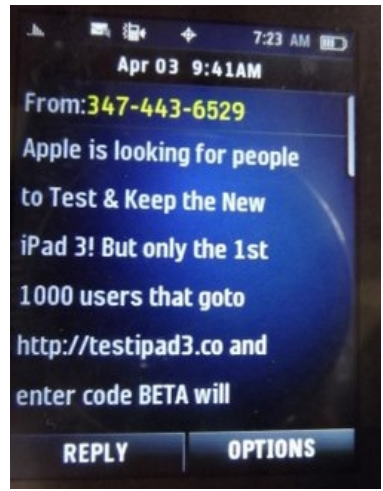

Figure 9. SMISHING

The website http://testipad3.co was registered by GoDaddy.com and was taken down the next day. The attack vector in this case was a reward. The message corresponds with the release of Apple's Ipad III.

## 5.3 Craigslist

Scams have scams on both the buyer and seller sides of a transaction. Both types of scams require breaking Craigslist's recommended policy of not completing a transaction in person.

### 5.3.1 Buyer Side Transaction

A trending fraud on the seller side of Craigslist transactions is for house rentals. A phisher will post an ad for house rental on Craigslist and will even include pictures of the house and give and address. The phisher "hooks" the phish, or victim, by offering an incredibly low rent on the house. The victim responds quickly to the offer to lock in the deal. The person offering the rental asks the victim for a security deposit to ensure that the victim gets the rental. The deal is never made and the victim never gets their deposit back.
> **Author's note:** *This exact scam happened to the step-sister of one of the authors.*

### 5.3.2 Seller Side Transaction
In a seller side phishing scam, the victim is the seller. A prospected buyer typically offers more than a seller is asking for on a deal, but the catch is that the buyer wants the item shipped. Below is a sample of a Craigslist scam. This particular scam shows some classic psychology of the Phisher. This message not only offers more money, but is appealing to the seller and the seller's trust by stating that the person is doing relief work in Haiti.

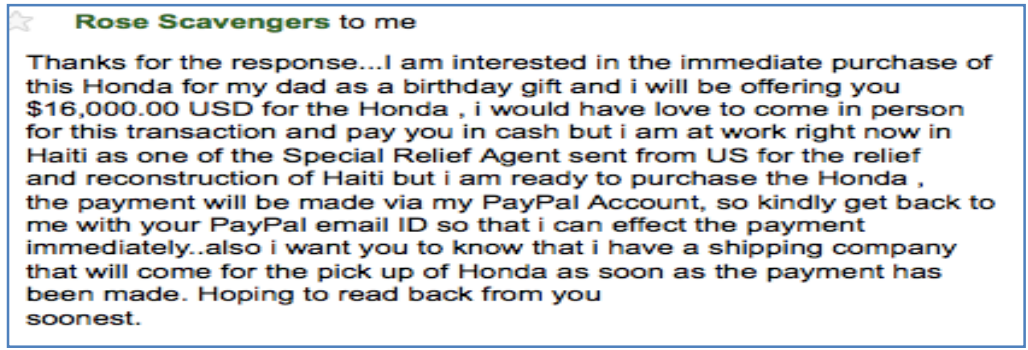

Figure 10. Craigslist Example [24]

## 6. RISK REDUCTION AND MITIGATION

The risks of Phishing or any other information security vulnerability can be reduced or mitigated, but never eliminated. Risk reduction begins at the senior management level of an organization.

Placing appropriate emphasis on IT security at the senior management and board of director's level is the first step toward minimizing system breaches. By establishing effective policies and procedures, boards of directors can promote an atmosphere that addresses critical security areas and establishes appropriate guidelines and standards for all employees. [25]

Senior management needs to have buy-in on security programs and fully fund them.

### 6.1 Training

Training is a key ingredient to making people aware of phishing scams. In addition to an awareness program, there is an online tool available through Carnegie Melon University and WombatSecurity.com. It is called Phishing Phil. It is a flash game that incorporates recognizing many of the attack vectors in this paper in a fun way. Phishing Phil requires a license, but is worth the investment.

### 6.2 Trojaned Hosts

The best defense against Trojans is to actively scan the machine for viruses with an up-to-date virus scanner. Virus scanners will catch most Trojans that are not zero-day exploits or custom variants. A virus scanner is only as good as its definitions. A zero-day exploit is an exploit that has not been identified or patched. Anti-virus programs rely on signatures for detection and if a Trojan is new there will not be a signature in anti-virus database. Trojans also exploit vulnerabilities in operating systems and application such as Java, Adobe Reader, and Flash. Keeping the operating system and applications at their most recent patch level reduces the exposure level of the system.

### 6.3 Man-in-the-Middle Attacks
#### 6.3.1 Transparent Proxy Servers
Transparent Proxy servers are difficult to detect. There are some third party proxy detectors available. It is good practice to employ a proxy detector even if phishing is not suspected since the main purpose of a proxy is circumvention of firewall rules.

International Journal of Network Security & Its Applications (IJNSA), Vol.5, No.4, July 2013

**6.3.2 DNS Cache Poisoning**

The first line of defense against DNS Cache Poisoning in an enterprise is ensuring that your DNS servers are properly configured and at a current patch level. Microsoft Servers offer DNS Protection for what Microsoft calls DNS polluting. "By default, Microsoft DNS servers, using Windows 2000 Service Pack 3 or later, acting as a parent in a child-parent relationship will fully perform cache pollution protection" [26].

Netcraft also offers a tool for detecting DNS poisoning. It is called the Netcraft toolbar. The Netcraft tool bar works by conducting a geographic lookup of the destination IP and displaying it on the webpage see below.

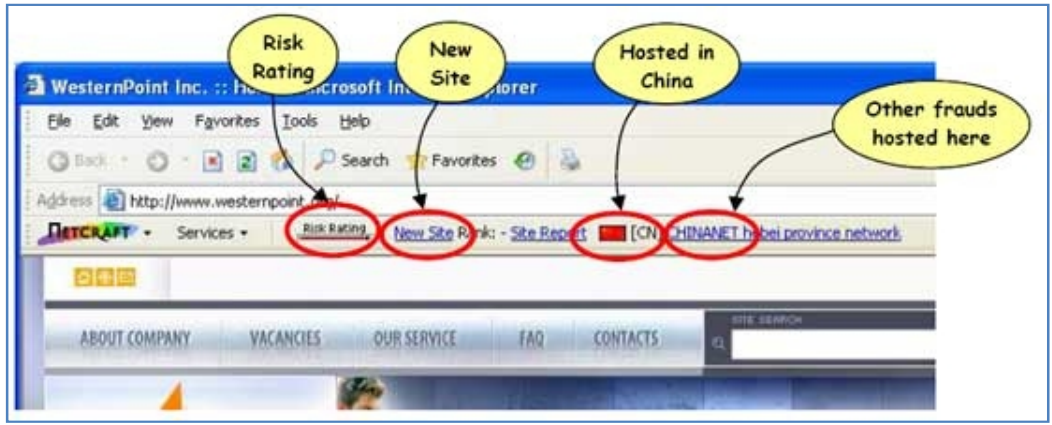

Figure 11.Netcraft

The Netcraft toolbar shows the risk rating of the website, the age of the website, the host location and whether or not there were ever frauds detected on the same site. Netcraft describes the functionality as "if your local DNS cache was poisoned such that the US Bank web site (http://www.usbank.com) pointed to an IP address located in Russia, then the toolbar would report the site as being located in Russia" [28].

**6.3.3 Browser Proxy Configuration**

In an enterprise environment where computers are in a client/server relationship in a domain, the permissions for browsers proxy should default to prohibited by the group policy. This adds a layer of protection and makes it difficult for an attacker to add a proxy since they would have to also override policy. For machines that are not part of a domain, a user should inspect the browser setting at a set interval.

**6.4 URL Obfuscation**
**6.4.1 Bad Domain Names**

Mitigating bad domain names via URL Obfuscation is accomplished by not trusting URLs that are provided. Users should not follow provided links rather type them in to the address or search bar themselves.

**6.4.2 Third Party Shortened URLs**
Third parties that offer URL shortening are aware of the misuse of their products and have added functionality. TinyURL.com has a preview feature that discloses the redirected site. Figure 12.Uses the early example of www.tinyurl.com/bankofamericafraudcomplaints.

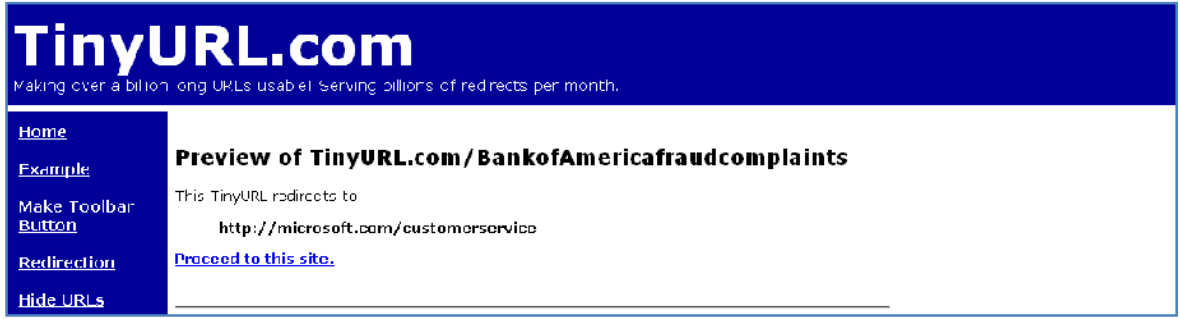

Figure 12. Tiny URL

The preview feature is enacted by appending the original link with the word "preview". An organization should train its employees to always preview any third-party shortened URLs. Using the example, the URL is: www.preview.tinyurl.com/bankofamericafraudcomplaints.

### 6.4.3 Host Name Obfuscation

The defense against host name obfuscation is to set a policy to not use obfuscated host names. This is similar to never using bad domain names. A user should always type in the common FQDN.

### 6.5 Cross-Site Scripting Attacks

Preventing XSS attacks requires diligence when code is created. Special attention is needed to ensure that a webpage is not vulnerable. There are tools available that check web applications for these vulnerabilities in code. Unfortunately these tools are also available to criminals. Typically a coder will create a filter that does not allow certain words to be added to a comment such as "<script>". This is a simple example on coding to prevent XSS. Actual rules and methods of prevention are outside the scope of this paper. Another way to mitigate the XSS is to have a penetration team routinely attempt XSS injection attacks. If your penetration can execute a XSS so can an attacker.

Having a well coded website is great, but it does not stop the common user from being exploited by the XSS codes. On the client side there are steps that can be taken to limit the effects of XSS. The first is through policy. Most organizations have a security policy limiting the use of the internet. Next is to ensure all computers are properly patched at the Operating System level and the Application level. Lastly, browsers can be configured to limit the ability to execute scripts on website. This however plays with the balance of productivity and security.

### 7. CONCLUSION

A solid awareness program keeps IA professionals and SAs abreast of techniques of phishers by being informative and relevant to classic and emerging attack vectors and methodologies to mitigate the risks. Good SAs keep the balance between security and productivity by knowing what is dangerous to the organization and what is not. This paper discussed the history of phishing from ancient Troy to AOL of the 1990s, Classic Phishing attack Vectors, Emerging Attack Vectors and processes to mitigate and limit the risks of phishing. The important aspect of phishing training is keeping the audience engaged. Providing historical examples of social engineering and phishing combined with technical examples of common attacks enforces the retention of the training materials.

[52] Rahman, Syed (Shawon) "System Security Specifications for a Multi-disciplinary Research Project",7th International Workshop on Software Engineering for Secure Systems conjunction with The 33rd IEEE/ACM International Conference on Software Engineering (ICSE 2011), May 21-28, 2011, Honolulu, Hawaii.

[53] Rahman, Syed (Shawon) and Donahue, Shannon; "Converging Physical and Information Security Risk Management", Executive Action Series, The Conference Board, Inc. 845 Third Avenue, New York, New York 10022-6679, United States

[54] Rahman, Syed (Shawon) and Peterson, Mike; "Security Specifications for a Multi-disciplinary Research Project"; The 2011 International Conference on Software Engineering Research and Practice (SERP'11), Las Vegas, Nevada, USA July 18-21, 2011

[55] Jungck, Kathleen and Rahman, Syed (Shawon); " Information Security Policy Concerns as Case Law Shifts toward Balance betyouen Employer Security and Employee Privacy"; The 2011 International Conference on Security and Management (SAM 2011), Las Vegas, Nevada, USA July 18-21, 2011

**Authors Bio:**

**Marc Rader** is a CISSP earning his B.S. degree in Information System Management from Park University in 2009 and his M.S. degree in Information Technology: Information Assurance and Security from Capella University in 2012.  Currently, he is pursuing his PhD in Information Technology: Information Assurance at Capella University.  Mr. Rader works as an Associate Faculty Member at Cochise College in Arizona and Network Engineer for TASC Inc. 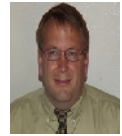

**Dr.  Syed (Shawon) M. Rahman** is an Associate professor in the Department of Computer Science and Engineering at the University of Hawaii-Hilo, Hawaii, USA and a part-time faculty of School of Business and Information Technology at the Capella University, Minneapolis, MN. Dr. Rahman's research interests include software engineering education, data visualization, information assurance a nd security, digital forensics, cloud computing security,  web accessibility, software testing and quality assurance. He has published more than 90 peer-reviewed articles. He is a member of many professionalorganizations including IEEE, ACM, ASEE, ASQ, ISACA, and UPE. 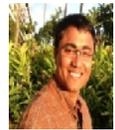